\documentstyle[12pt,aasms4]{article}
\lefthead{Aparicio {\it et al.}}
\righthead{The SFH of Pegasus DIG}

\begin{document}

\title{The Star Formation History of the Pegasus Dwarf Irregular Galaxy\\}

\author{A. Aparicio}
\affil{Instituto de Astrof\'\i sica de Canarias, E38200 - La Laguna, Tenerife, 
Canary Islands, Spain}

\author{C. Gallart}
\affil{Carnegie Observatories, 813 Santa Barbara  St., Pasadena, CA 91101, USA}

\and

\author{G. Bertelli}
\affil{Cosiglio delle Ricerche, Vicolo dell'Osservatorio 5, I35122 - Padova, 
Italy}

\begin{abstract}

The star formation history (SFH) of the Pegasus dIr, a likely Local Group member 
at 0.95 Mpc from the Milky Way, is investigated. We characterize the SFH by two 
basic functions: the star formation rate, $\psi(t)$, and the chemical 
enrichment law, $Z(t)$. It has been derived by comparing the color-magnitude 
diagram of the  resolved stars in Pegasus, with a total of 189 model diagrams produced with different $\psi(t)$ and $Z(t)$ laws.

The models in better agreement with the data indicate that star formation 
began in Pegasus about 15 Gyr ago and was larger, on average, during the first half than during the second half of the galaxy's life. During the most recent epoch, for which the SFH can be obtained with much 
better time resolution,  the star formation seems to be produced in a bursting mode. This may 
have been the case for the whole life of the galaxy, although the resolution in time towards 
older epochs is not good enough to actually detect it. As for the chemical enrichment law, 
the best way to account for the observed metallicity of the galaxy 
($Z_f=0.002^{+0.002}_{-0.001}$) is that it suffered a prompt initial chemical 
enrichment. This would be the case if infall was important, at least during the primeval epoch of galaxy evolution and points to a picture in which the galaxy began 
forming stars and enriching its interstellar medium in an early phase of 
collaps, when a lot of gas had still to be added to it.

Pegasus and NGC~6822 are the only dIrs for which the kind of analysis presented here has been done. The fact that, like Pegasus, NGC~6822 also shows an 
important old to intermediate-age stellar population indicates that the Baade's 
sheet observed in most dIr, may in fact be the signature of an important 
population of old stars and suggest that dIr actually are old objects populated 
by large numbers of old stars.

The mass in stars and stellar remnants is derived from average of the best model SFHs obtained. The percentage of dark matter in Pegasus that cannot be accounted for with stellar remnants or with an extrapolation of the Kroupa {\it et al.} IMF down to $\sim 0.1$ M$_\odot$ turns out to be $\sim 92\%$. 

\end{abstract}

\section{Introduction \label{introduction}}

Nearby galaxies offer the opportunity of studying their stellar content directly 
from the color-magnitude (CM) diagram even when observed with ground-based 
telescopes. This is the most direct way of deriving the star formation history 
(SFH) of such systems, which is in turn a basic step toward the knowledge of the physical 
properties of galaxy evolution. We asume the SFH to be a function of time accounting for all the characteristics which 
determine the formation of stars, mainly the star formation rate (SFR), the 
chemical enrichment law (CEL) and the initial mass function (IMF).

Several studies of this kind have been done 
for dwarf irregular (dIr) galaxies. Some representative papers
are by Aparicio {\it et al.}\markcite{sexa} (1987), Bresolin, Capaccioli \& Piotto 
\markcite{bresolin} (1993), Marconi {\it et al.}\markcite{marconi} (1995), Tolstoy 
\markcite{tolsleoa} (1997), Gallart {\it et al.} \markcite{n68sfh1} (1996b), among 
others. The last three works use synthetic CM 
diagrams to derive the distribution of stellar ages and then the SFH (Bertelli 
{\it et al.} \markcite{bertlmc} 1992 is an early example of the use of synthetic 
diagrams to derive the SFH of the LMC). This is a powerful technique for obtaining the SFH in these galaxies.
However, most of the existing studies of dIrs limit the analysis to the 
last few hundred million years of the galaxy's history, which is presumably a 
very small fraction of the total lifetime of the galaxy. Gallart et 
al.\markcite{n68sfh1} (1996b,\markcite{n68sfh2}c), is a first example of how the SFH, extending to old and intermediate ages, of a dIr (NGC~6822) can be analyzed using  synthetic diagrams 
together with a careful simulation of crowding effects. $V$ 
and $I$ photometry of evolved low mass
stars is used to get information on the corresponding evolutionary phases on 
the CM diagram. In this way, it is possible to obtain an important insight into the SFHs in galaxies, and extending them towards old ages, even if key structures such as the main sequence (MS) turn-off, the horizontal branch (HB) or red clump (RC) of the helium-burning phase are not 
observed. The time resolution of the SFH worsens 
for older ages but its general trend (including an estimate of the age for the 
beginning of the star formation) can still be obtained. The reason for the 
loss of resolution in the SFH is that stars older than $\sim$ 1 Gyr clump 
together in the reddest part of the CM diagram during their evolution on the red 
giant branch (RGB) and asymptotic giant branch (AGB), forming two structures 
that we term the {\it red-tangle} and {\it red-tail} (see Aparicio \& Gallart 
\markcite{chile} 1994, and Gallart {\it et al.} \markcite{n68dat} 1996a). A very 
enlightening figure showing the distribution of stars younger and older than 1 
Gyr is given in Fig.~1 of Gallart {\it et al.} (1996b). 

In this paper, the results for the SFH of the Pegasus dIr, obtained using the method outlined above, are presented. Pegasus (UGC~12613, DDO~216) was discovered by Holmberg \markcite{holmberg} (1958) and was studied for the first time using a CCD by Hoessel \& Mould \markcite{peghoe1} (1982). The Cepheid estimate for the distance to Pegasus (1.7 Mpc, Hoessel {\it et al.} \markcite{peghoe2} 1993) is in disagreement by a factor of two with the tip of the RGB (TRGB) estimate (0.95 Mpc, Aparicio \markcite{pegdist} 1994). In this paper, we will use the TRGB estimate, 0.95 Mpc. Aparicio \& Gallart \markcite{peg1} (1995, hereafter Paper~I) present the 
photometry of the resolved stars in Pegasus used here as well as a detailed 
discussion of crowding effects and the method for simulating these effects in 
synthetic CM diagrams. More information about this method can be found in 
Gallart {\it et al.} (1996b).

Pegasus shows few traces of ionized gas. To our present knowledge, Pegasus has just two small HII regions and some difuse emission. The total H$\alpha$ luminosity has been reported by Hunter, Hawley \& Gallagher~\markcite{hunter}(1993). For the distance assumed here it turns out that $L_{\rm H\alpha}=2.7\times 10^{36}$ erg\,s$^{-1}$. The brightest HII region has a total H$\alpha$ luminosity of $(3.4\pm 0.1)\times 10^{35}$ erg\,s$^{-1}$ (Paper~I). This luminosity can be produced by a 
single main sequence B0 star. Skillman, Bomans \& Kobulnicky \markcite{pegz} 
(1997) studied this same HII region and found a new, weaker one. They agree that a single main sequence B0 star is enough to excite this bright HII region and 
found an oxygen abundance O/H$=(8.4\pm 2.5)\times 10^{-5}$ which corresponds to 
$Z=0.002^{+0.002}_{-0.001}$ (assuming (O/H)$_{\odot}=8.3\times 10^{-4}$ 
and $Z_\odot=0.02$). This abundance will be used to constraint the final metallicity in the SFH models. 

In Sec.~\ref{problems} the problems and methods for 
determining the SFH are discussed. The different assumptions made in order to determine the SFH and the results obtained for Pegasus are presented in Sec.~\ref{quantitative}. In Sec.~\ref{parameters}, these results are discussed and some integrated parameters are provided. In Sec.~\ref{conclusions} the main conclusions of the work are summarized.

\section{Derivation of the star formation history: problems and methods \label{problems}}

Provided it is deep enough, the $[(V-I),I]$ CM diagram of an LG dIr galaxy contains stars with ages\footnote{In the following discussion we call {\it young} stars those younger than 1 Gyr; {\it intermediate-age} stars those with ages in the range 1--10 Gyr and {\it old} stars those older than 10 Gyr.} within the whole range covered by the SFH of the 
galaxy. Young stars are found in the MS, bright blue-loops 
(BL) and red supergiant (RSG) strip, while intermediate-age and old stars are 
mixed in the red-tangle and the red-tail (see Aparicio \& 
Gallart \markcite{chile} 1994, and Gallart {\it et al.} 1996a for a thorough 
discussion). The information contained in the red-tail and the red-tangle raises the possibility of obtaining some information on the SFH, from the beginning of the galaxy's activity as a star forming system to the present time. 

The MS is the only place where a derivation of the SFR as a function of time would be formally simple: if we assume a known IMF, the distribution of stellar luminosities along the MS would drive to the distribution of ages. In practice, however, the MS can provide information only for the last few Myr of the galaxy's history. Moreover, at the distances typical for LG dIrs ($\sim 1$ Mpc) and using ground-based observations, the MS appears mixed with the BL strip and is too noisy to provide detailed information about the time dependence of the SFR. Nevertheless, the distribution of stars along the BL and RSG strips can be used to complete the information necessary for obtaining the SFR as a function of time for young ages. 

Stars older than $\sim 1$ Gyr (i.e., intermediate-age and old stars) populate the red-tangle and the red-tail. This last is formed in principle by AGB stars only. Since this is a short-lived phase, the red-tail is expected to be 
populated by a relatively small number of stars. The red-tangle contains 
a major fraction of the stars in typical CM diagrams and provides most of the 
information on the old and intermediate-age SFH. Nevertheless, the intermixing 
of stars over almost the whole interval of possible ages and metallicities in the red-tangle makes it difficult to infer the SFH from it. In particular, the well known age-metallicity degeneracy in the RGB plays its role in making up the tangle. (see Fig.~1 of Aparicio \& Gallart 1994, and Fig.~3 of Freedman~\markcite{freedman}1994).
This degeneracy, and the fact that evolved stars of all ages end their life in this small area of the CM diagram are the reasons why isochrone fitting type techniques are not successful in interpreting this kind of data. Nevertheless, when stellar evolution theory is used to actually derive a distribution of stars in the CM diagram, taking into account additional information or reasonable assumptions (such as the current metallicity of the system or the fact that $Z(t)$ is expected to increase with time), and a realistic simulation of the errors is performed, it is possible to overcome the limitations of the data to a certain degree, and get important information on the past SFH of the system.
Using model CM diagrams is a useful method for this task, although a worsening of the time resolution for old ages cannot be avoided. The 
method is based on the comparison of the distribution of 
stars in the observed CM diagram with those of a set of model diagrams built 
with different input SFHs. The model diagram best reproducing the observed one 
provides the sought after SFH. In practice, several models will be compatible with the 
observations. The solution will, in general, not be unique but, considered 
together, all valid models will show the general trend of the SFH 
including the age of the beginning of star formation, and will give an idea of the errors.

Construction of a model CM diagram consists of two steps: ({\it i}) computation 
of a synthetic CM diagram, and ({\it ii}) simulation of observational effects. 
Details of the computation of synthetic CM diagrams are discussed in Gallart 
{\it et al.} (1996b), whereas the procedure for simulating observational effects on them, as well as the nature of these effects, is explained in Paper~I and in Gallart {\it et al.} (1996a,b). In short, synthetic CM diagrams are based on a set of stellar evolutionary tracks covering a wide range of metallicities and masses. The Padua stellar evolution library is used here (see Bertelli {\it et al.}\markcite{tracks} 1994, and references therein). The two key functions we use to define the SFH are the SFR and the CEL, both functions of time. We 
will denote these functions as $\psi(t)$ and $Z(t)$, respectively. These are the 
free inputs that are changed to generate the set of model CM diagrams. The IMF 
is also an input which we choose to be fixed. In particular, the 
IMF derived by Kroupa, Tout \& Gilmore \markcite{imf} 
(1993) for the solar neighborhood has been used. Synthetic CM diagrams give the 
distribution of stars for different input SFHs, 
but they cannot be directly compared with observed CM diagrams, because the latter 
include observational effects, mainly resulting from crowding. These effects 
are of three kinds: ({\it i}) a fraction of stars are lost, ({\it ii}) magnitudes 
and color indices are systematically shifted, and ({\it iii}) they are affected by large external 
errors. The three effects are strong, non-trivial functions of the magnitude and 
color index of each star and of the distribution of magnitudes and color indices 
of all the stars present in the galaxy (see Paper~I, and Gallart {\it et al.} 1996a). 

The procedure of simulating observational effects in synthetic CM diagrams consists of two steps. First, a table is constructed with the injected and 
recovered magnitudes of a large sample of artificial stars added to the original image in several steps (DAOPHOT~II is used for this task; see Stetson \markcite{dao} 1993); lost artificial stars are also included in the table with a flag indicating that 
they have not been recovered. We call this table the {\it crowding trial table}. 
Secondly, for each star in the synthetic CM diagram the following procedure is 
performed. Let $m_s$ and $c_s$ be the magnitude and color of the synthetic star. 
An artificial star with initial (injected) magnitude $m_i$ and color $c_i$ in a 
given interval around $m_s$ and $c_s$ is randomly picked out from the crowding 
table. If it has been lost, the synthetic star is removed from the CM diagram. 
If it has been recovered with magnitude $m_r$ and color $c_r$, the differences 
$\delta m=m_r-m_i$, $\delta c=c_r-c_i$ are added to the magnitude and color of 
the synthetic star to obtain its final values as $m_s+\delta m$ and $c_s+\delta 
c$. The reader is referred to the above cited papers for a thorough discussion 
of this procedure. Suffice it to remark here that this method has the advantage that the simulation of observational effects is performed using direct information from the artificial stars trials and no assumptions are made about either the nature of the effects or the propagation of errors and completeness factors. The only hypothesis is that crowding effects on the artificial stars are a good representation of crowding effects on the observed real stars. The only requirement necessary to make this hypothesis 
reasonable is that the distribution of color indices of synthetic and real 
star samples must be similar. This is automatically satisfied in our case 
because we are in fact trying to reproduce the distribution of magnitudes and color indices of the real star sample, hence we must work with a similar distribution for 
the synthetic stars.

In the following subsections, the observed and models CM diagrams used for analyzing the Pegasus SFH are presented.

\subsection{Observational CM diagram \label{observational}}

The observational data of Pegasus used here have been 
discussed in Paper~I. Figure~\ref{obs} shows the [$(V-I)_0$,$M_I$] 
CM diagram. A distance modulus of $(m-M)_0=24.9$ and a reddening of 
$E(V-I)=0.03$ have been used after Aparicio (1994). The different features present in the CM diagram are labeled. For a qualitative 
description of them, the reader is referred to Paper~I.

\placefigure{obs} 

\subsection{Model CM diagrams \label{models}}

Following the procedure outlined above, we produce a set of model CM diagrams 
for varying $\psi(t)$ and $Z(t)$, using in all cases the crowding trial table obtained in Paper~I. Results of simulations using similar procedures as those used here can be seen in Gallart {\it et al.} (1996b) and in Aparicio {\it et al.}\markcite{hst} (1996). From here on, we will refer only to these final model CM diagrams and in this section we will focus our attention on the discussion of the modeling we have done of the SFH. Together with the shapes of $\psi(t)$ and $Z(t)$, the initial and final time values ($T_i$ and $T_f$) and metallicity ($Z_i=Z(T_i)$ and $Z_f=Z(T_f)$) are basic input parameters. As we stated above, the IMF by Kroupa {\it et al.} (1993) is used as a fixed input.

\subsubsection{Chemical enrichment law \label{cel}}

The chemical enrichment law, $Z(t)$, is one of the two functions we are using to characterize the SFH. Choosing such-and-such a $Z(t)$ may have important consequences on the resulting $\psi(t)$. But chemical enrichment processes in galaxies are very uncertain and we have no independent information on what is the most reliable $Z(t)$ law. For this reason we have 
used three different approaches to $Z(t)$ that cover a wide range of enrichment scenarios. In this way, we are testing each particular $\psi(t)$ for three quite extreme possibilities of $Z(t)$, trying to avoid the bias that would be introduced in the selection of $\psi(t)$ if only one possibility for $Z(t)$ were considered, as a consequence of the age-metallicity degeneracy in the RGB: if the $Z(t)$ chosen had been {\it too slow}, it would have favoured $older$ $\psi(t)$, and vice versa if $Z(t)$ were {\it too fast}. 

We produced three sets of chemical laws, which we will denote by C$_{\rm A}$, C$_{\rm B}$ and C$_{\rm C}$:

\begin{itemize}
\item C$_{\rm A}$: a logarithmic function of the form 
$$Z(t)=Z_i+k\,\ln\,\mu(t)^{-1}$$ 
This relation is valid for a galaxy in which an arbitrary fraction of the material returned to the 
interstellar medium after stellar evolution escapes from the system. A particular case of this is a closed system, which is produced when $k$ equals the yield (see Tinsley \markcite{tins} 1980). Nothing is assumed {\it a priori} about $k$. It is determined after $Z_i$, $Z_f$, $\mu_i$ and $\mu_f$ have 
been chosen. $\mu(t)$ is the fraction of gas present in the galaxy at time $t$ considering only the mass that participates in the chemical evolution process $\mu(t)=M_g(t)/[M_g(t)+M_\star(t)]$. Note that in this set, $Z(t)$ implicitly depends, through $\mu(t)$, on the $\psi(t)$ function at work in each model. $Z_i=0.0001$ and two different values for $Z_f$ ($Z_f=0.002$ and 0.004) have been used. Two more values of $Z_f$ ($Z_f=0.006$ and 0.008) have been tested, but the results are not considered here because these $Z_f$ are not compatible with the value observed by Skillman {\it et al.}~(1997). Some comments are given in Sec.~\ref{general}.

Initial and final values $\mu_i=1$ and $\mu_f=0.2$ have been used for the 
models. The latter is an estimate of the present gas fraction of Pegasus using 
data of $L_B$ and $M_H$ from Lo {\it et al.}~\markcite{lo}(1993) and 
$M_\star/L_B\lesssim 1$ estimated from Larson \& Tinsley's \markcite{larson} 
(1978) models. After our model diagrams were computed, Hoffman \markcite{hoff} 
{\it et al.} (1996) published new HI data for Pegasus. Using these data, a better 
estimate of the final gas fraction is $\mu_f\simeq 0.39$ (see Sec.~\ref{parameters}). This would have produced a higher metallicity during the whole 
galaxy's life, except at the fixed initial and final times, so that  
resulting models would be intermediate between these and those of case C$_{\rm 
B}$ (see below).

\item C$_{\rm B}$: a linear function of the integral of $\psi(t)$ of the form 
$$Z(t)=Z_i+k\,I_{\psi}(t)$$ 
where $I_{\psi}(t)=\int_{T_i}^t\psi(t^\prime)dt^\prime$. The same 
values for $Z_i$ and $Z_f$ as in set C$_{\rm A}$ have been used, including in addition, the value $Z_f=0.001$. This form represents an enrichment faster 
than that of case C$_{\rm A}$, which is expected in systems with moderate infall.

\item C$_{\rm C}$: a linear function of time, regardless of $\psi(t)$, with 
$Z_i=0.002$ and $Z_f=0.002$ or 0.004. This simulates a prompt initial chemical 
enrichment followed by a shallow 
(or even null) enrichment during the rest of the galaxy's life, as would be 
the case for a system with large infall rate in which the initial mass of 
unenriched gas is small compared with the SFR. Models with initial or final metallicity lower than $Z=0.002$ have 
not been checked in this case because even models with $Z_i=Z_f=0.002$ fail to 
produce good results, because they have too low an overall metallicity (as indicated by the fact that they produce red-tangles which are too blue; see below).

\end{itemize}
\subsubsection{Star formation rate \label{sfr}}

We have divided the analysis of the SFR into two parts: we have first investigated the general trend of $\psi(t)$ for the whole lifetime of the galaxy, and then we have refined its shape for the last 0.4 Gyr to obtain a best representation of the distribution of stars in the regions of the diagram populated by young stars. In this way, we benefit from the better resolution in the temporal sampling that the CM diagram offers for stars younger than a few hundred Myr. The procedure used for determining the young SFH is simpler than that for the general SFH and is explained in Sec.~\ref{young}.

For the general form, we have defined 27 different shapes for $\psi(t)$ (see 
Fig.~\ref{dibujito}). These shapes involve different values of $T_i$ and $T_f$ 
and follow the criterion that $\psi(t)$ is a step function which can take three possible values (0, 1 and 4) on an arbitrary scale which will be later transformed into an absolute scale of M$_\odot$yr$^{-1}$ or M$_\odot$yr$^{-1}$pc$^{-2}$. The 27 different shapes are defined by the times at which $\psi(t)$ changes. These times can be 15, 12, 9, 6, 3, 1, and 0 Gyr on a scale where $t=0$ is the present time and $t$ increases toward the past. For practical purposes the star formation is stopped at 10 Myr ago rather than at $t=0$. The 27 shapes are schematically shown in Fig.~\ref{dibujito}. We have organized them in four sets, which we will term S$_{\rm A}$, S$_{\rm B}$, S$_{\rm C}$ and S$_{\rm D}$, respectively. Models are grouped in each set depending on whether the star formation has been mainly produced in a burst (S$_{\rm C}$, S$_{\rm D}$) or continously (S$_{\rm A}$, S$_{\rm B}$) and whether an underlying star formation is (S$_{\rm B}$, S$_{\rm D}$) or is not (S$_{\rm A}$, S$_{\rm C}$) allowed. The different shapes are ordered inside each 
set following a progressive scale of {\it youth} of the global stellar 
population. Following these sequences, there are some shapes that can be 
included in more than one set (e.g., number 1 of sets S$_{\rm A}$ and S$_{\rm 
C}$). These shapes have been repeated where necessary in Fig.~\ref{dibujito}
to show a more comprehensive picture. From now on, we will use a code to designate each particular model when necessary. The code has the form C$_\alpha n$S$_\beta m$, where C$_\alpha$ refers to the CEL set; $n$ stands for the final 
metallicity; S$_\beta$ refers to the SFR set and $m$ is the number of the 
$\psi(t)$ shape, as shown in Fig.~\ref{dibujito}. As an example, $\rm 
C_B4S_A3$ is the model with $Z(t)$ from set C$_{\rm B}$ having $Z_f=0.004$ and 
$\psi(t)$ of the form number 3 of set $\rm S_A$ of Fig.~\ref{dibujito}.

\placefigure{dibujito}

It should be noted that $Z(t)$ of C$_{\rm 
C}$ are independent of $\psi(t)$; i.e. for each final metallicity of set C$_{\rm C}$, the metallicity at a given age is always the same. But for the other two cases, $Z(t)$ depends of the specific $\psi(t)$ used, hence each of the 27 former shapes has its own associated sets of $Z(t)$ functions for C$_{\rm A}$ and C$_{\rm B}$.

\section {Quantitative calculation of the star formation history \label{quantitative}}
\subsection {The general star formation history \label{general}}

As stated in Sec.~\ref{problems}, the method we follow for analyzing the general trend of the SFH is based in the comparison of the distribution of stars in the observed CM diagram with those of a set of model diagrams built with different input SFHs. In the following subsections we will shortly present the indicators used to decide what models adequately reproduce the observed CM diagram and what have been the results obtained.

\subsubsection {Indicators and their errors \label{indicators}}

The comparison of the distribution of stars between model and observed CM 
diagrams is made through a number of indicators sensitive to the distribution of stellar ages and metallicities and hence on the SFH of the galaxy. They represent the relative number and distribution of stars in the red-tangle and the red-tail, as well as the position and shape of those features. We have used a total of nine indicators. Six of them are relative to the position and shape of the red-tangle as well as to the distribution of stars in it. The remaining three indicators are relative to the same in the red-tail as well as to the ratio between red-tail and red-tangle stars. All the indicators are defined in Apendix~A. Some of them are similar to the ones used in Gallart {\it et al.} (1996b). The reader is referred to that paper for further details. 

All the indicators, as computed from the model CM diagrams, are affected by stochastic errors produced in the algorithms for generating the synthetic CM diagram and for simulating crowding effects. To estimate these errors we have generated 20 model CM diagrams with a particular set of inputs and the same number of stars used in the other models, but changing the seed of the aleatory numbers generated in the algorithms. We have determined all the indicators for these 20 models, and adopted the corresponding $\sigma$ values as the random error for each one. It has been assumed that the error of each indicator is the same for all the model diagrams. 

Indicators as computed from the observed CM diagram are affected by other kind of errors: those coming from the uncertainties in the reddening and the 
distance modulus. To estimate their effects, we have calculated the indicators 
for the observed CM diagram shifting it by all the possible combinations of 
extreme values of $E(V-I)$ and $(m-M)_I$ inside the intervals provided by the 
adopted observational errors ($\pm 0.02$ and $\pm 0.1$, respectively). The errors of the indicators are calculated taking into account the extreme values obtained about their central values. 

There is of course another important source of errors: the unknowns of stellar evolutionary theory, but these are difficult to estimate explicitly (see Sec.~\ref{uncertainty}). To minimize their effects, wide intervals in $\sigma$ will be used to consider model diagrams indicators compatible with those of the observed diagram. 

\subsubsection {Results for the general star formation history \label{results}}

The 27 forms for $\psi(t)$ (Fig.~\ref{dibujito}) discussed above were 
computed with each of the seven possible CELs grouped in sets C$_{\rm A}$, 
C$_{\rm B}$ and C$_{\rm C}$. This produced a total of 189 models which were 
compared using the indicators defined in Appendix~A with the Pegasus CM diagram. 

Pegasus shows evidence of recent star formation activity. Nevertheless, models stopping the star formation several Gyr ago should not be rejected for this  reason alone, because some star formation in the last few Myr can always be added without changing the characteristics of the red-tail and the red-tangle, which are the structures used in this section to derive the general trend of the SFH at intermediate-age and old epochs. Details of the recent star formation history will be considered in Sec.~\ref{young}.      

A broad criterion to select models that would be compatible with the observed CM diagram of Pegasus has been used. Rejected models will have very little chance of representing Pegasus, and so we will know with high confidence what the SFH of Pegasus {\it is not} like. The selected models, considered all together, will show the general trend of the Pegasus SFH. The criterion has been to select the models which have the six red-tangle indicators not further away than 3$\sigma$ from the interval allowed by observations, and the three indicators involving red-tail stars not further away than 5$\sigma$ from the observational interval. A wider interval is allowed for red-tail indicators because we are less confident in AGB stellar evolutionary models. Note that it is necessary that all the indicators be satisfied in order to have a good overall agreement between model and observed CM diagrams. For example, the fact that the red-tangle of a model CM diagram is significantly shifted from the observed one is a reason enough to reject the model, no matter of how well shaped the red-tangle might be.

The ten models meeting this criterion are (see Fig.~\ref{dibujito} and Sec.~\ref{sfr}) $\rm C_B4S_A2$, $\rm C_C4S_A5$, $\rm C_C4S_A6$, $\rm C_C4S_B1$, $\rm C_C4S_B2$, $\rm C_C4S_B3$, $\rm C_C4S_B4$, $\rm C_C4S_B5$, $\rm C_C4S_D2$, and $\rm C_C4S_D3$. 

The selected models can be used to sketch the general characteristics of Pegasus' SFH:

\begin{itemize}
\item[1.] The important common characteristics of the ten accepted models are that, independently of the $Z(t)$ law, they show star formation starting 15 Gyr ago and at a higher rate, on average, during the first half than during the second half of the galaxy's life. 

\item[2.] The models with low initial 
metallicity ($Z_i=0.0001$: sets C$_{\rm A}$ and C$_{\rm B}$) produce red-tangles that are too blue and are in general rejected. The only exception is model $\rm 
C_B4S_A2$. 

\item[3.] As for models with a prompt initial chemical enrichment ($Z_i=0.002$: set C$_{\rm C}$) all those with $Z_f=0.002$ are rejected, but 9 with $Z_f=0.004$ are accepted. The fact that $Z_f=0.002$ is not allowed is a consequence of the position (too blue) of the red-tangle. In the accepted models, there is a tendency for more stars to have been formed in the first half of galaxy's life than in the second half. But in general, star formation activity at a low or high rate, maintained until the present time, is required. The only exception is ($\rm C_C4S_A5$), which stopped star formation only 1~Gyr ago. 

\end{itemize}

Consequently, the emerging scenario is that Pegasus seems to have started forming stars $\sim15$ Gyr ago and shows a large fraction of old stars. As for the $Z(t)$ law, it seems that metallicity has evolved quickly at the beginning of the history of the galaxy, in such a way that a very small number of stars would have metallicities lower than $Z=0.002$. Nevertheless, a certain amount of metallicity dispersion at all ages is reasonably expected. 

As mentioned above, besides the 189 models used for the former analysis, several models with $Z_f=0.006$ or 0.008 have also been tested for sets $\rm C_A$ and $\rm C_B$. Several of them passed the former criteria. All of them have star formation starting 15 Gyr ago, and most show a $\psi(t)$ enhanced for the first half of the galaxy's life. These models satisfactorily reproduce the Pegasus SFH, but they are not considered here because their final metallicities are not compatible with the observed value $Z=0.002^{+0.002}_{-0.001}$.

The combination of all the models compatible with observations provides a good indication of how the SFH should have been and, at the same time, of the limitations that we have in this determination of the SFH, due to relatively little information contained in the data. Observation of the horizontal branch and red clump, and still more, possible intermediate-age main sequence turn-offs would allow a characterization of some details of the SFH, whereas now we are only delineating its general trend. 

\placefigure{10best}

Figure~\ref{10best} shows the $\psi(t)$ functions of the ten accepted models overimposed on each other. Normalization has been done using the number of stars in the upper part of each model and Pegasus' red-tangles and taking into account that, for Pegasus' distance (0.95~Mpc), we are covering a field of $\sim 663\times 663$ pc$^2$. Each single $\psi(t)$ has been drawn using the same density of dots. In this way, the larger the density of dots in a given area, the larger is the number of models common to that particular area of the $[t,\psi(t)]$ plane. In other words, a high density of dots at a given $\psi(t)$ indicates that most models show, {\it at least} that level of $\psi(t)$. Figure~\ref{10best} must consequently be interpreted as follows: a given density of dots at a given time is related to the probability of Pegasus having a star formation rate larger than that shown by the density of dots itself. For example, it is very likely that $\psi(t)>0.2\times 10^{-3}$ M$_\odot$yr$^{-1}$ for any time. It is also possible that $\psi(t)>10^{-3}$ M$_\odot$yr$^{-1}$ for $t>8$ Gyr, but that is unlike for $t<5$ Gyr (or at least none of the selected models shows such a high $\psi(t)$ for that recent epoch). To complete the information given in Fig.~\ref{10best}, the $\psi(t)$ function averaged for all the accepted models is also shown. Error bars are the $\sigma$ of the sample. Together with the values of $\psi(t)$, Fig.~\ref{10best} shows the two most important results that we have mentioned above: the star formation very likely started $\sim15$ Gyr ago and is probably a decreasing function.

Although  $\psi(t)$ is quite uncertain, its temporal integral is much better defined. The average SFR can be obtained using the information from all the accepted models as 
$$\bar\psi={\sum^n_{i=1}{1\over T}\int\psi_i(t)dt\over n}$$
\noindent where $i$ refers to each accepted model, $n=10$ is the number of accepted models and the integral extends to the whole galaxy's life time, $T=15$ Gyr. It results 
$$\bar\psi=(6.7\pm 0.4)\times 10^{-4} {\rm M}_\odot{\rm yr}^{-1}$$
or
$$\bar\psi=(1.5\pm 0.1)\times 10^{-4} {\rm M}_\odot{\rm yr}^{-1}{\rm pc}^{-2}$$

\subsection {The young star formation history \label{young}}

Contrary to what happens to the old and intermediate-age stars, 
which are completely mixed into the red-tail and red-tangle, the 
distribution of young stars in the CM diagram has a clear dependence on age.
Therefore, the SFR $\psi(t)$ can be determined in more detail for the last 
several hundred Myr using the information provided by the distribution of stars 
in the blue part of the CM diagram and in the regions corresponding to the RSG 
and the bright AGBs. For this further analysis we have defined six regions in 
the CM diagram, all of them bluer than $(V-I)\leq0.8$ and two more in the red, 
brightest part of the diagram, as indicated in Fig.~\ref{8reg}.

\placefigure{8reg}
\placefigure{edades}

The dependence on age of the relative amount of stars in each of the regions of 
Fig.~\ref{8reg} is illustrated in Fig.~\ref{edades}, which shows 
the distribution of star counts that a constant $\psi(t)$ (during the last 0.5 
Gyr at least) would produce across the eight regions for each of five 
logarithmic age intervals. To plot this figure, a model CM diagram with a large 
number of stars (some 2200 inside the eight defined regions) has been computed 
with constant SFR from $5\times10^8$ yr ago to the present time (we will term it $\psi_0(t)$). Each of the five distributions has been normalized to the 
total number of stars in the eight regions. In this way, Fig.~\ref{edades} 
represents in what extent each $\log\,age$ interval is represented in the CM diagram and where the stars in a given $\log\,age$ interval are preferentially 
placed in that diagram. It must be noted that region~8 corresponds to bright AGBs. The star counts in this region would potentially be very useful. Unfortunately, as we have previously discussed, the evolution of intermediate-mass AGBs is poorly understood and the number of stars in this region will be considered only indicatively.

\placefigure{8counts}

Figure~\ref{8counts} shows the star counts for the eight regions for the constant $\psi_0(t)$ initial model CM diagram (dashed line) compared with the same star counts for Pegasus (solid line). The model data have been arbitrarily normalized to put them on a scale similar to that for Pegasus. This figure gives a first insight into the actual shape of the recent $\psi(t)$. It is clear that the constant SFR, $\psi_0(t)$, produces too many stars in 
regions~2 and 5. The overpopulation of region~2 indicates that $\psi(t)$ has been relatively lower in 
the interval $\log\,t=7.0-8.0$. The overpopulation of region~5 indicates that 
$\psi(t)$ has been low in the interval $\log\,t=8.0-8.4$. These are just qualitative 
insights, since the final result for $\psi(t)$ must account for the counts in 
all the regions. The fact that the distribution of stars in the eight regions defined in the CM diagram is clearly dependent on age allows a simple, accurate way to derive the 
recent $\psi(t)$: starting at the distributions of star counts as a function of 
$\log\,age$ for each region for the model CM diagram built up with $\psi_0(t)$ 
SFR (Fig.~\ref{edades}), an iterative procedure is performed to determine the fraction of stars in each age interval that have to be selected from the model to reproduce the number of stars found for Pegasus in each region. Eight $\log\,age$ intervals have been defined from $\log\,age=7.0$ to 8.6, in steps of 0.2. The simplification that $\psi(t)$ is a step function changing only at these times has been imposed. To plot Fig.~\ref{edades}, and, for simplicity, the five $\log\,age$ intervals used from $\log\,age=7.0$ to 8.0 have been grouped into only two intervals.

The number of stars in each region of the CM diagram must satisfy the following eight relations:
$$N_r=\sum_{i=1}^8a_iF_i(r)$$
\noindent where $N_r$ is the number of stars in region $r$ ($r$ varies from 1 to 
8), $F_i(r)$ is the number of stars from the $i$-th $\log\,age$ interval that 
populate region $r$ for a constant $\psi_0(t)$ and $a_i$ is the fraction of stars conserved in the $i$-th age interval. But the former eight relations cannot be satisfied exactly since an algebraic solution produce negative values for $a_i$. A simple trial-and-error procedure has been followed to solve the system approximately. Multiplication of the $a_i$ parameters by $\psi_0(t)$ directly yields the relative SFR for each time interval and hence $\psi(t)$. The resulting function is shown in Fig.~\ref{psifin_1}. Normalization has been done comparing the total number of blue stars (stars bluer than $(V-I)_0\leq 0.8$) in the models and in the Pegasus CM diagram. Errors are difficult to estimate in such a trial-and-error procedure. Checking different values for the $a_i$ coeficients, we estimate that errors in $\psi(t)$ are about 30\%. 
The distribution of stars in the eight regions of the CM diagram defined in Fig.~\ref{8reg} produced by this $\psi(t)$ is also shown in Fig.~\ref{8counts} as a dotted line.

\placefigure{psifin_1}

All our model diagrams are built up to $t=10$ Myr; {\it i.e.}, they formally do not give information for the last 10 Myr of the galaxy's history. As a consequence, the $\rm H\alpha$ flux of the galaxy has been used for estimating $\psi(t)$ for $t\leq 0.01$ Gyr. It results from the analysis based on the model diagrams, that $\psi(t)=3.4\times 10^{-4}$ M$_\odot$yr$^{-1}$ at $t=10$ Myr. Extrapolating this value to the present and using the Kroupa {\it et al.} (1993) IMF, it turns out that Pegasus should contain $\sim 5$ stars more massive than 15 M$_\odot$. But the H$\alpha$ flux of the brightest HII region (Aparicio \& Gallart 1995; Skillman {\it et al.} 1997) indicates that there is only one such star and that therefore, $\psi(t)$ should be smaller in the last 10 Myr 
than at $t=10$ Myr. $\psi(t)\sim (1.0\pm 0.3)\times 10^{-4}$ M$_\odot$yr$^{-1}$ 
for the last 10 Myr would account for between one and two live stars more 
massive than 15 M$_\odot$. Normalizing to the covered area, it turns out that  $\psi(t)\sim (2.3\pm 0.7)\times 10^{-10}$ M$_\odot$yr$^{-1}$pc$^{-2}$.

\subsection{Sources of uncertainty \label{uncertainty}}

The errors most difficult to 
evaluate are those produced by the unknowns of stellar evolutionary theory. These 
can produce features in the CM diagram appearing in positions and with shapes 
slightly different from the observed ones. Short-lived, low-populated stellar 
evolutionary phases produce further errors due to random fluctuations in small numbers. But the simple solution of assuming errors to be related to the square root of the 
star counts in each region is not necessarily valid, since  predictions in the number of 
stars which should populate a given area of the diagram can not be considered 
less significant when that number is small. Both effects (shifting of features and random fluctuations) are minimized because the different regions we have used 
in our analysis have been defined in such a way as to include, as far as possible, 
stars in the desired evolutionary phases only. This procedure may introduce 
some subjectivity into the analysis, but it avoids a more undesirable effect: if 
model and observed CM diagrams were compared using entirely objective methods 
({\it e.g.}, maximum likelihood, $\chi^2$ or bidimensional Kolmogorov), the results could be largely affected by systematic errors in the stellar evolutionary models. 

Among the existing unsolved problems in stellar evolution, the modeling of bright AGBs is particularly uncertain. Mass loss plays an 
important role in this phase, but its modeling is not completely understood, 
and bolometric corrections for the reddest AGBs are not well established. The 
first produces overpopulation or underpopulation of bright AGBs, depending on 
the mass-loss model. The second makes the bright AGBs have colors quite 
different from those predicted by the theory. For these reasons, bright AGBs have received less weight in our analysis than other stars.

A different kind of uncertainty is that of the temporal resolution of the 
derived SFH. As we have already stated, this worsens for older ages but, again, it is difficult to determine explicitly. A rough qualitative insight can be 
obtained by considering the common characteristics of accepted models. Fig.~\ref{10best} is an attempt of showing the common trends of those models. The fact that this figure appears as a blurred distribution of dots is indicative of uncertainty. It is particularly interesting to note (Fig.~\ref{psifin_1}) that, over the last 400 Myr, where the time resolution is greater, the star formation appears as an episodic phenomenon which occurs at time intervals of several 10 Myr. It could be the case that star formation has always been that way, but we can not detect these short-term changes for older ages. For the youngest history itself, our resolution is not better than a few 10 Myr in the best case and we can not exclude the possibility that the star formation turns on and off over shorter intervals of time, although our results are consistent with a constant $\psi(t)$ from 10 to 60 Myr. 

\section {Final global results and integrated parameters \label{parameters}}

Results shown in Fig.~\ref{10best} and \ref{psifin_1} can be combined to plot a full representation of Pegasus' $\psi(t)$, from 15 Gyr ago to the present time. The result is shown in Fig.~\ref{psifin_2}, where error bars for the last 0.4 Gyr are not plotted for clarity. For a complete representation of the Pegasus SFH, $Z(t)$ has to be included. A useful display is the population box introduced by Hodge (1989). Figure~\ref{popbox} is the Pegasus population box. The horizontal axis is 
time, the vertical axis is $\psi(t)$ and the third axis is $Z(t)$. The $Z(t)$ represented is the one noted as $\rm C_C4$. This is the law correponding to nine of the ten accepted models.

\placefigure{psifin_2}
\placefigure{popbox}

Figure~\ref{sinfin} shows the model CM diagram produced by a combination of the ten accepted general models and the result for the young $\psi(t)$ obtained in Sec.~\ref{young}. This combination corresponds very approximately to the SFH shown in Fig.~\ref{popbox}. Figure~\ref{sinfin} fails to reproduce the number of bright AGB stars in Pegasus. These stars come from different young age intervals and the origin of the discrepancy is very likely in the stellar evolution models (see Sec.~\ref{uncertainty}). 

\placefigure{sinfin}
\placefigure{malos}

For comparison and illustrative purposes in the kind of model diagrams we are using, Fig.~\ref{malos} shows two rejected models: $\rm C_A2S_A6$ (panel A) and $\rm C_C4S_A9$ (panel B). Model $\rm C_A2S_A6$ corresponds to a constant $\psi(t)$ and a logarithmic $Z(t)$, starting at a very low value and finishing in $Z=0.002$, which is the observational value (Skillman {\it et al.}~1997). This diagram clearly ilustrates how such a metallicity law fails to reproduce the Pegasus CM diagram. Model $\rm C_C4S_A9$ has the same metallicity law as nine of the ten accepted models, but its $\psi(t)$ starts at 3 Gyr ago. This diagram shows how a stellar population lacking a significant number of old stars is unlikely in Pegasus. 

The average rate at which Pegasus forms stars is quite low. A time average of 
the SFR shown in Fig.~\ref{psifin_2} or \ref{popbox} results in 
$\bar{\psi}=6.7\times 10^{-4}$ M$_\odot$yr$^{-1}$ or $1.5\times 10^{-9}$ 
M$_\odot$yr$^{-1}$pc$^{-2}$. The fraction of matter due to stars and stellar 
remnants can also be calculated using the derived SFH and the assumed IMF. This 
mass turns out to be $M_\star =8\times 10^6$ M$_\odot$. This is a lower estimate, however, because the images we are using do not cover the whole galaxy. Taking into account that the surface brightness is larger in the central part of the galaxy,  and that we are covering $\sim 2/3$ of the galaxy's optical body (Aparicio \& Gallart 1995), a better estimate could be $M_\star\simeq 10^7$ M$_\odot$. Using data by Hoffman {\it et al.} (1996) and the distance to Pegasus of 0.95 Mpc from Aparicio (1994), the HI mass of Pegasus is $M_{\rm HI}=4.9\times 10^6$ M$_\odot$ and, multiplying by 4/3 to account for the He mass, $M_{\rm gas}=6.5\times 10^6$ M$_\odot$ hence the gas fraction relative to the total mass intervening in the chemical evolution is $\mu=M_{\rm gas}/(M_{\rm gas}+M_\star)=0.39$. Finally, using data from Hoffman {\it et al.}~(1996) corrected for a distance of 0.95 Mpc, the total dynamical mass of Pegasus is estimated to be $M_{\rm tot}=1.8\times 10^8$ M$_\odot$, which implies that $\sim 92\%$ of this total mass is produced by dark matter which is neither explained by stellar remnants nor by an extrapolation to low masses of the Kroupa {\it et al.} (1993) IMF. All these data, together with the luminosity (data are taken from de Vaucouleurs {\it et al.}~\markcite{dvau}1991) and mass-luminosity relations are summarized in Table~1. Here, $A$ is the area covered by our field ($663\times 663$ pc$^2$) and $\psi(0)$ refers to the present SFR. The last line contains the distance to the barycenter of the LG, considered to be in the 
line connecting M31 and the Milky Way, at a distance of 0.45 Mpc from the Milky 
Way. This results from adopting, after Peebles \markcite{peebles}~(1989), a mass 
for the Milky Way 0.7 times that of M31 and neglecting the masses of any other 
galaxy in the LG.

\section{Conclusions \label{conclusions}}

We have analyzed the SFH of the Pegasus dIr galaxy. Our study is based on the 
$[(V-I),I]$ CM diagram of the galaxy and 189 model CM diagrams computed with 
different input SFHs, considered to be represented by the $\psi(t)$ and $Z(t)$ functions, the IMF being a fixed input function. 
The model CM diagrams which best reproduce the observed one give the sought after SFH of the galaxy. There are four basic input parameters: the initial and final values of time ($T_i$ and $T_f$), and the initial and final values of metallicity ($Z_i$ and $Z_f$). Together with the 
shapes of $\psi(t)$ and $Z(t)$ themselves, these parameters complete the 
information necessary to define each model. The analysis of $\psi(t)$ has 
been divided into two stages. 
The reason for this is that the regions of the CM diagram where young stars and 
intermediate-age and old stars are distributed are different. First, we looked for a general $\psi(t)$ that is formally valid for 
intermediate and old ages (from $t=15$ Gyr to $t=1$ Gyr). Ten models satisfactorily reproduce the observed CM diagram. Their common properties give the trend of the Pegasus SFH. A more accurate answer for the last few hundred Myr is also provided, with better time resolution obtained from the distribution of blue stars, RSGs and bright AGBs in the CM diagram. Our models stop at $t=10$ Myr, but an estimate of $\psi(t)$ for $t<10$ Myr is provided by the H$\alpha$ flux of the galaxy.

The characteristics of the Pegasus SFH can be summarized as follows:

\begin{itemize}
\item Independently of the $Z(t)$ law, Pegasus probably started forming stars $\sim 15$ Gyr ago. On average, its SFR was larger in the first half than in the second half of the galaxy's life.
\item For the last few hundred Myr, where the time resolution is better, the star formation appears to be produced in a bursting mode; {\it i.e.}, short periods of 
enhanced SFR followed by short periods of low SFR. The time intervals in which 
these bursts are produced and their duration cannot be securely established 
from our analysis because we may not have enough time resolution. Nevertheless, 
our resolution in time is good enough to say that the SFR has been low and 
roughly constant for the last 10 to 60 Myr.
\item Pegasus seems to have had a prompt initial chemical enrichment. In accordance with usual chemical evolutionary scenarios (see for example 
Peimbert {\it et al.} 1993), this would imply an important infall, at least during the 
primeval epoch, and points to a picture in which the galaxy began to form stars 
and enrich the interstellar medium at an early phase of its collapse, when a large fraction of the gas had still to be added to the galaxy.
\end{itemize}

We have shown that star formation very likely began $\sim 15$ Gyr ago in Pegasus and that it was relatively high in the past. This is a similar case to that of NGC~6822 (Gallart {\it et al.} 1996b). These are the only two dIr galaxies for which the kind of analysis of the old SFH presented here, giving a quantitative estimate of the SFR towards early epochs, has been performed. The fact that in both cases an important population of old to intermediate-age stars has been found supports the idea that dIrs are old objects with a large amount of old stars, and that the so-called Baade sheet (Baade \markcite{baade} 1963) is likely the signature of an old population.

The average rate at which Pegasus is currently forming stars is quite low, about 
$\bar{\psi}=1.5\times 10^{-9}$ M$_\odot$yr$^{-1}$pc$^{-2}$. The mass 
of stars and stellar remnants is estimated to be about $M_\star\simeq 
10^7$ M$_\odot$ and the gas fraction relative to the total mass intervening in 
the chemical evolution  $\mu=M_{\rm gas}/(M_{\rm gas}+M_\star)=0.39$. Finally $\sim 92\%$ 
of the total mass of Pegasus is produced by dark matter which is neither 
explained by stellar remnants nor by an extrapolation to low masses of the 
Kroupa {\it et al.}~(1993) IMF.

\newpage
\appendix 

\section {Definition of indicators \label{appendix}}
\subsection {Indicators relative to the red-tangle}

Six indicators have been used to set the 
shape and position of the red-tangle in the CM diagram and the distribution of stars inside it. Unless stated otherwise, only stars falling in the 
region limited by $-3.75<I<-2.25$ and by two lines defined by points 
$[(V-I),I]=[0,-1]$ and $[1,-4]$ and $[(V-I),I]=[1.5,-1]$ and 
$[2,-4]$ are considered. The stellar color indices are used 
first to compute a color function, the CID1 defined in Gallart et 
al. (1996b). This distribution is then used in the calculation of the 
indicators. These are:

\begin{itemize}
\item {\bf EDG-R}, representing the red edge of the red-tangle at $I\simeq -3.4$. 
In practice it is defined as the $(V-I)$ color which leaves bluewards 95\% of 
the red-tangle's stars with magnitudes in the interval $-3.75\leq I<-3.0$. 
\item {\bf EDG-B}, representing the blue edge of the red-tangle at $I\simeq 
-3.75$. In practice it is defined as the $(V-I)$ which leaves redwards 95\% of  
red-tangle stars with magnitudes in the interval $-3.9\leq I<-3.6$ (this is the only 
red-tangle-related indicator computed using stars brighter than $I=-3.75$). 
\item {\bf MAX}, representing the color index of the center of the red-tangle at 
$I\simeq-3.4$. It is calculated as the median of the red-tangle 
distribution of stellar color for stars in the interval $-3.75\leq I<-3.0$.
\item {\bf FWHM}, the full width at half of maximum of the red-tangle 
distribution of stellar colors at $I\simeq-3.4$. Stars in the interval 
$-3.75\leq I<-3.0$ have been used.
\item {\bf TUD}, the rate between the number of red-tangle stars brighter 
and fainter than $I=-3.0$. Stars in the interval $-3.75\leq I<-2.25$ have been 
used.
\item {\bf RRB}, the rate from redder to bluer stars in the red-tangle. For 
this, each red-tangle star is considered {\it redder} or {\it bluer} if 
it lies redwards or bluewards with respect to the line defined by the points 
$[(V-I),I]=[1.3,-1]$ and $[1.55,-4]$. Stars bluer than $(V-I)=1.1$ are not 
considered in order to avoid contamination from young blue-loop stars. Only stars in the 
interval $-3.75\leq I<-2.25$ have been used.
\end{itemize}

\subsection {Indicators relative to the red-tail}

Three indicators have been used to set the position and extension of the red-tail and the ratio between stars in it and in the red-tangle. Stars are 
considered as red-tail members if 
they have magnitudes in the interval $-5.75<I<-3.75$ and are placed redwards with
respect to the line defined by the points $[(V-I),I]=[1.5,-4]$ and $[2.0,-8]$. 
To avoid the inclusion of stars at the TRGB, stars bluer than $(V-I)=2.0$ are considered 
as red-tail stars only if they are brighter than $I=-4.25$. The luminosity 
and color distribution functions of red-tail
stars, LFAGB and CID2 (see Gallart {\it et al.} 1996b) have been calculated and the 
following three indicators have been then defined:

\begin{itemize}
\item {\bf 70M}: Given the luminosity function of AGB stars (LFAGB) computed in the 
color interval $2.4\leq(V-I)<3.0$, 70M is the $I$ magnitude leaving upwards 
(brighter magnitudes) 70\% of that distribution.
\item {\bf 70C}: Given the color distribution in the red-tail (CID2), 70C is the 
$(V-I)$ color leaving blueward 70\% of that distribution.
\item {\bf RTT}: ratio of red-tail to red-tangle stars. It gives information about intermediate-age to old $\psi(t)$, as well as about $Z(t)$.
\end{itemize}

\newpage
\figcaption[obs.eps]{Observational CM diagram of Pegasus. The different stellar evolutionary phases present in the CM diagram are labeled. The horizontal and vertical tracks joining at $[(V-I),I]=[2,-4]$ have been plotted for easy comparison with diagrams in Figs.~\ref{sinfin} and \ref{malos}.
\label{obs}}
\figcaption[dibujito.eps]{Schematic representation of the $\psi(t)$ shapes used 
to generate the model CM diagrams. They are divided into four sets depending on 
wether the star formation has been mainly produced in an extended way (S$_{\rm A}$, S$_{\rm B}$) or in a burst (S$_{\rm C}$, S$_{\rm D}$) and whether an
underlying star formation is (S$_{\rm B}$, S$_{\rm D}$) or is not (S$_{\rm A}$, 
S$_{\rm C}$) allowed. Shapes are ordered 
inside each set following a sequence of {\it youth}. Some shapes are repeated in 
different sets in order to show a clearer sequence. 
\label{dibujito}}
\figcaption[10best.eps]{Representation of the ten general $\psi(t)$ accepted models. Density of dots in a given area, is related to the number of models that have in common that particular area of the $[t,\psi(t)]$ plane. Consequently, a given density of dots at a given time is related to the probability of Pegasus having a SFR larger than that shown by the density of dots itself. The $\psi(t)$ function averaged for all the accepted models is shown by a thick full line. Error bars are the $\sigma$ of the sample.
\label{10best}}
\figcaption[8reg.eps]{The eight regions used for the determination of the 
younger part of $\psi(t)$ superimposed on the Pegasus CM diagram.
\label{8reg}}
\figcaption[edades.eps]{Relative number of stars of different logarithmic age intervals which 
populate each of the eight regions shown in Fig.~\ref{8reg}. The logarithmic age 
intervals used are indicated in the upper part of the figure.
\label{edades}}
\figcaption[8counts.eps]{Number of stars populating each of the eight regions 
defined in Fig.~\ref{8reg}. The full line shows data for Pegasus. The dashed line 
shows the distribution of stars in a model CM diagram generated with constant 
$\psi_0(t)$. The dotted line is the distribution of stars in the model CM 
corresponding to the finally accepted $\psi(t)$ for the last 0.4 Gyr.
\label{8counts}}
\figcaption[psifin_1.eps]{Pegasus SFR for the last 0.6 Gyr
\label{psifin_1}}
\figcaption[psifin_2.eps]{The complete SFR for Pegasus obtained plotting together the general $\psi(t)$ shown in Fig.~\ref{10best} and the recent $\psi(t)$ shown in Fig.~\ref{psifin_1}.
\label{psifin_2}}
\figcaption[popbox.eps]{The population box of Pegasus representing the SFH 
of the galaxy combining the two key functions: $\psi(t)$ and $Z(t)$.
\label{popbox}}
\figcaption[sinfin.eps]{The model CM diagram generated by the SFH shown 
in Fig.~\ref{popbox}. The horizontal and vertical tracks joining at $[(V-I),I]=[2,-4]$ have been plotted for easy comparison with diagrams in Figs.~\ref{obs} and \ref{malos}.
\label{sinfin}}
\figcaption[malos.eps]{Two examples of model CM diagrams not compatible with Pegasus. The upper panel diagram corresponds to model $\rm C_A2S_A6$. The lower panel corresponds to model $\rm C_C4S_A9$. The horizontal and vertical tracks joining at $[(V-I),I]=[2,-4]$ have been plotted for easy comparison with diagrams in Figs.~\ref{obs} and \ref{sinfin}.
\label{malos}}

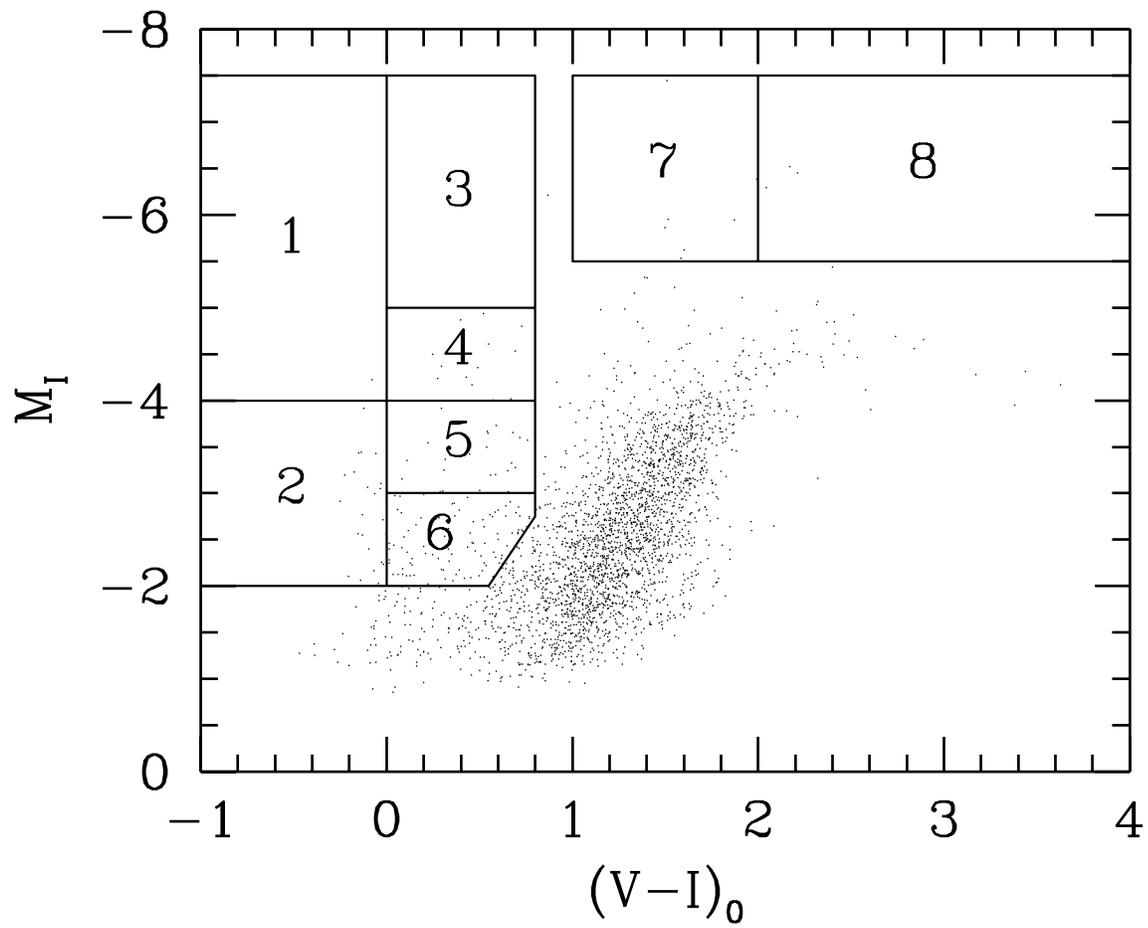
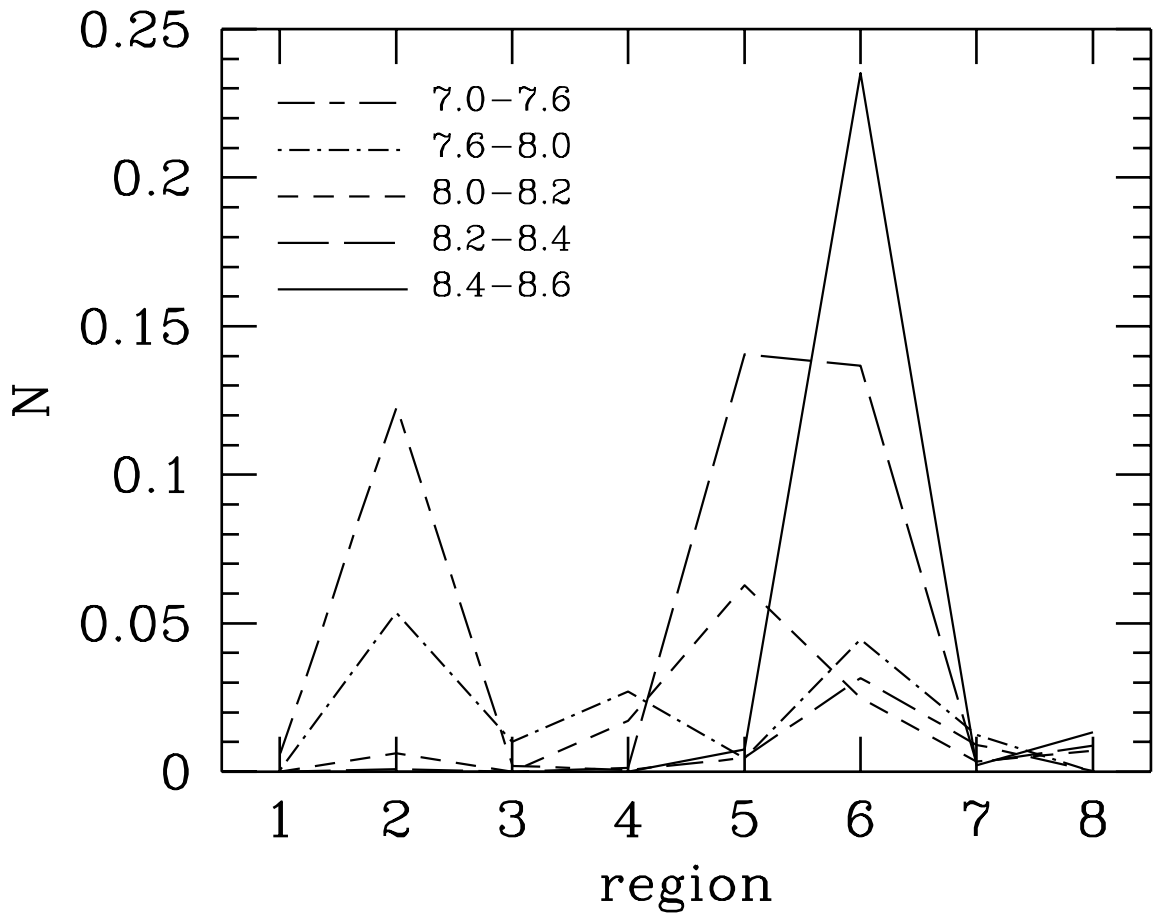
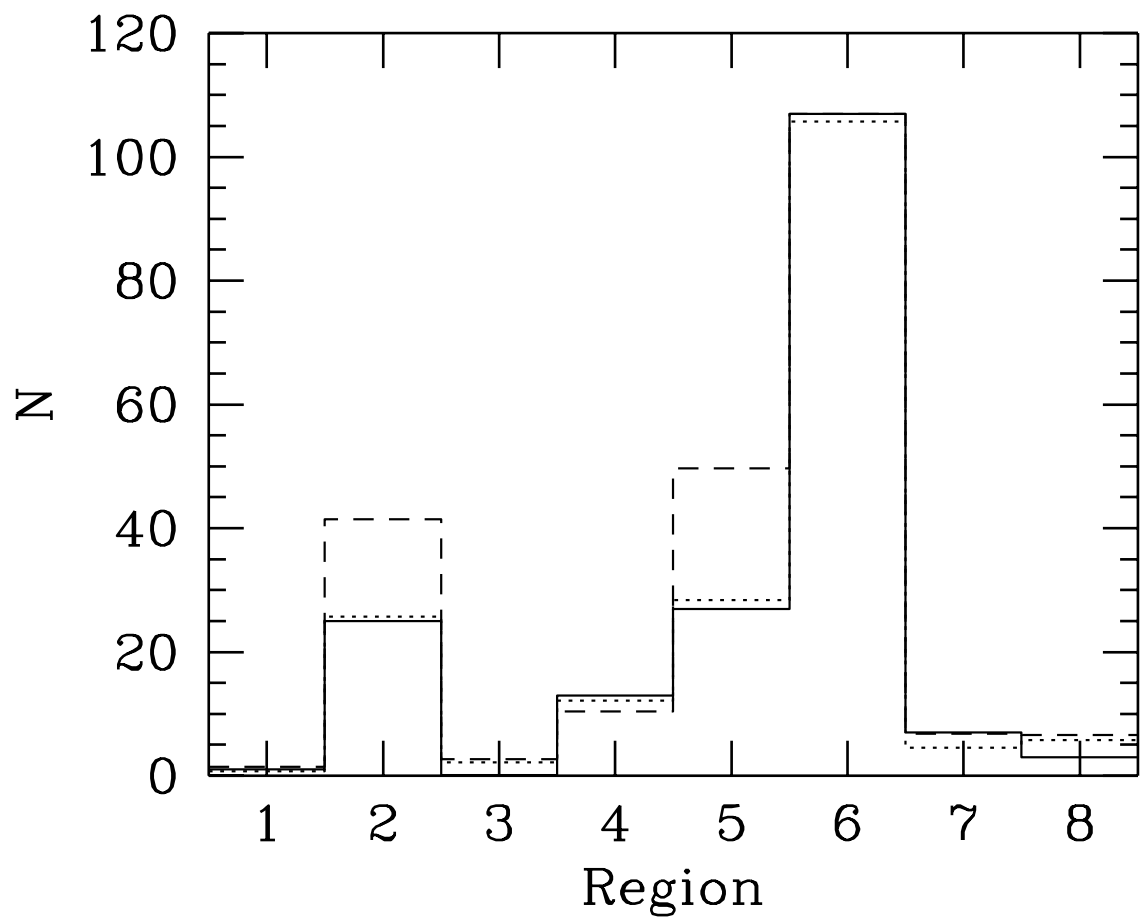
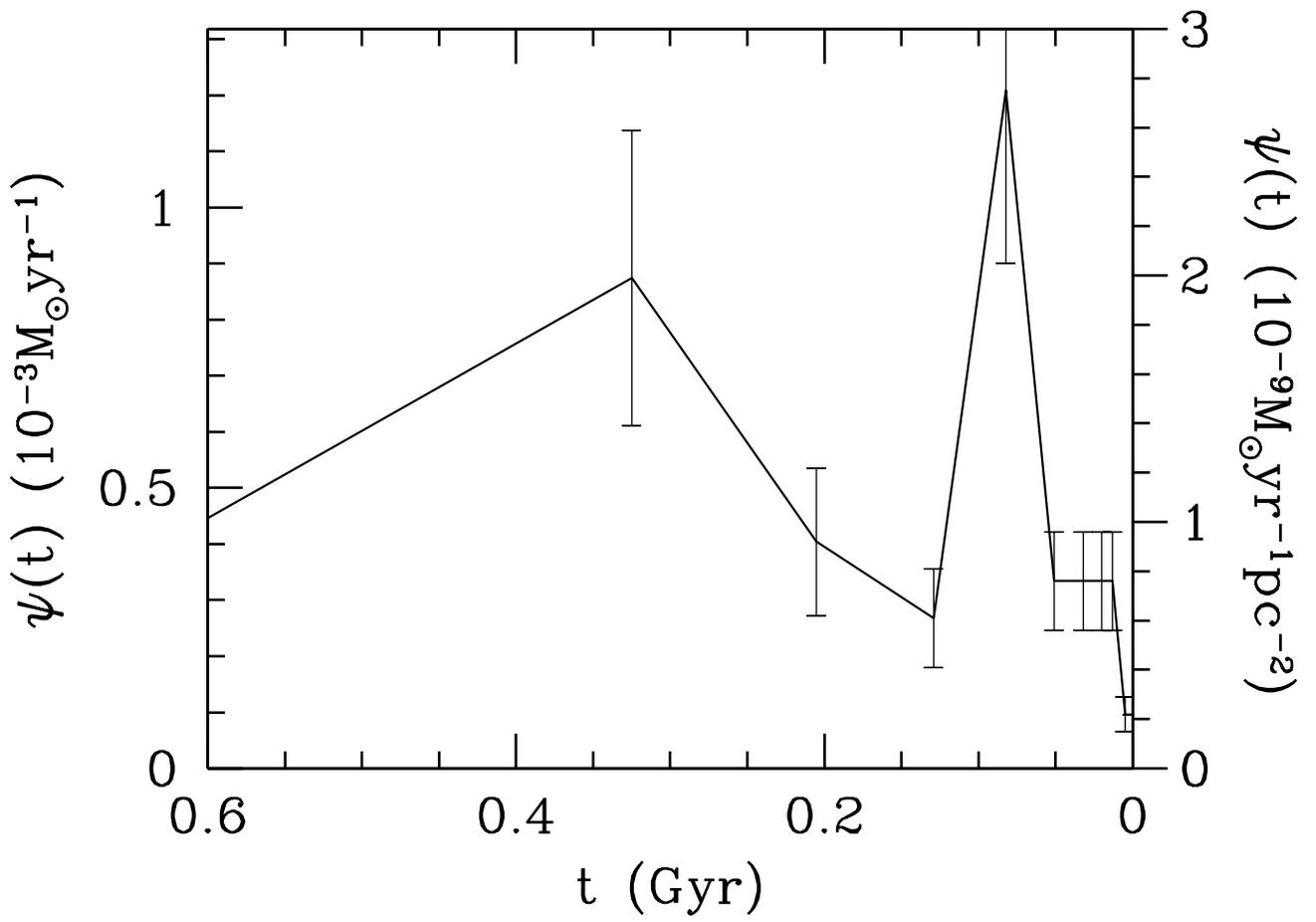
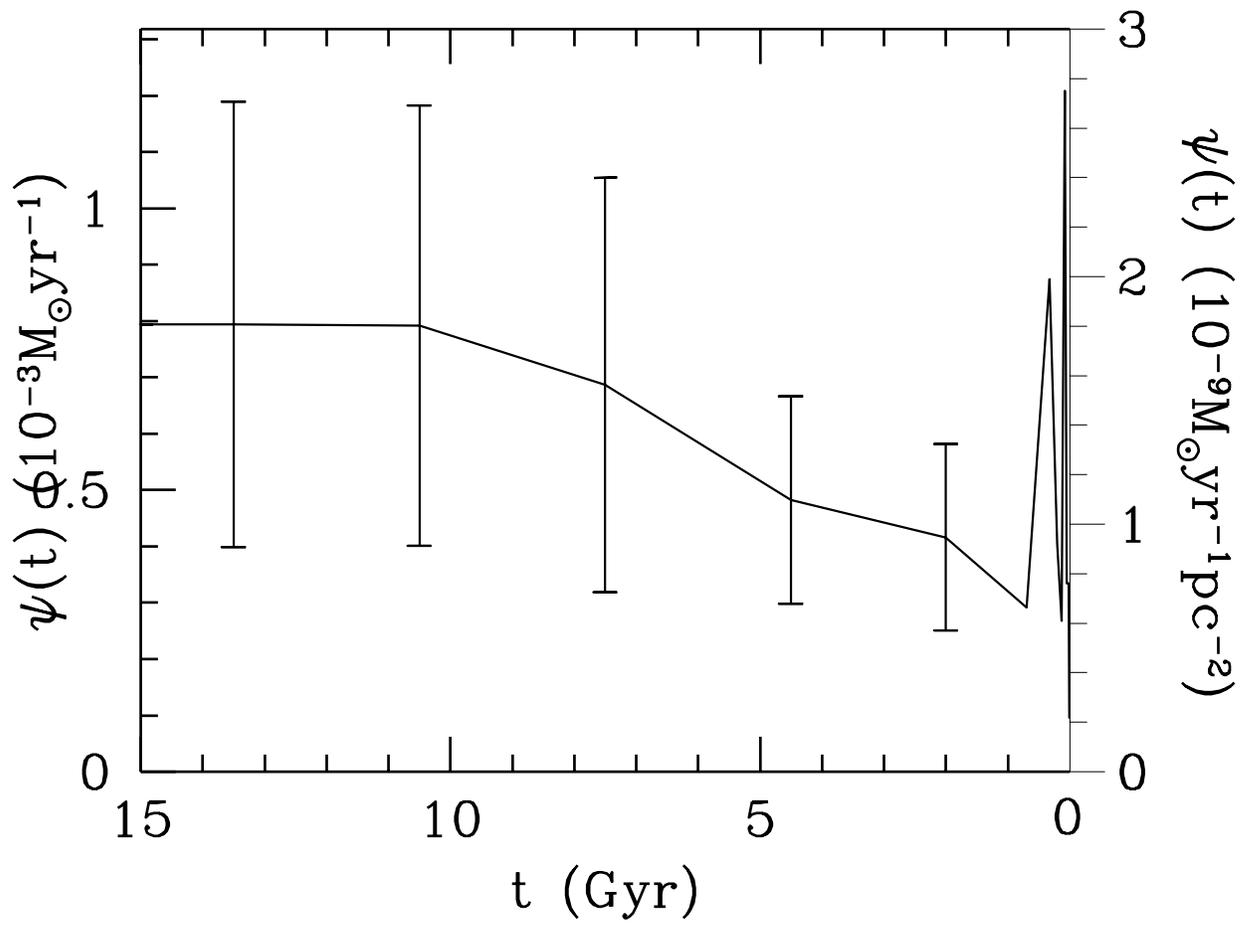
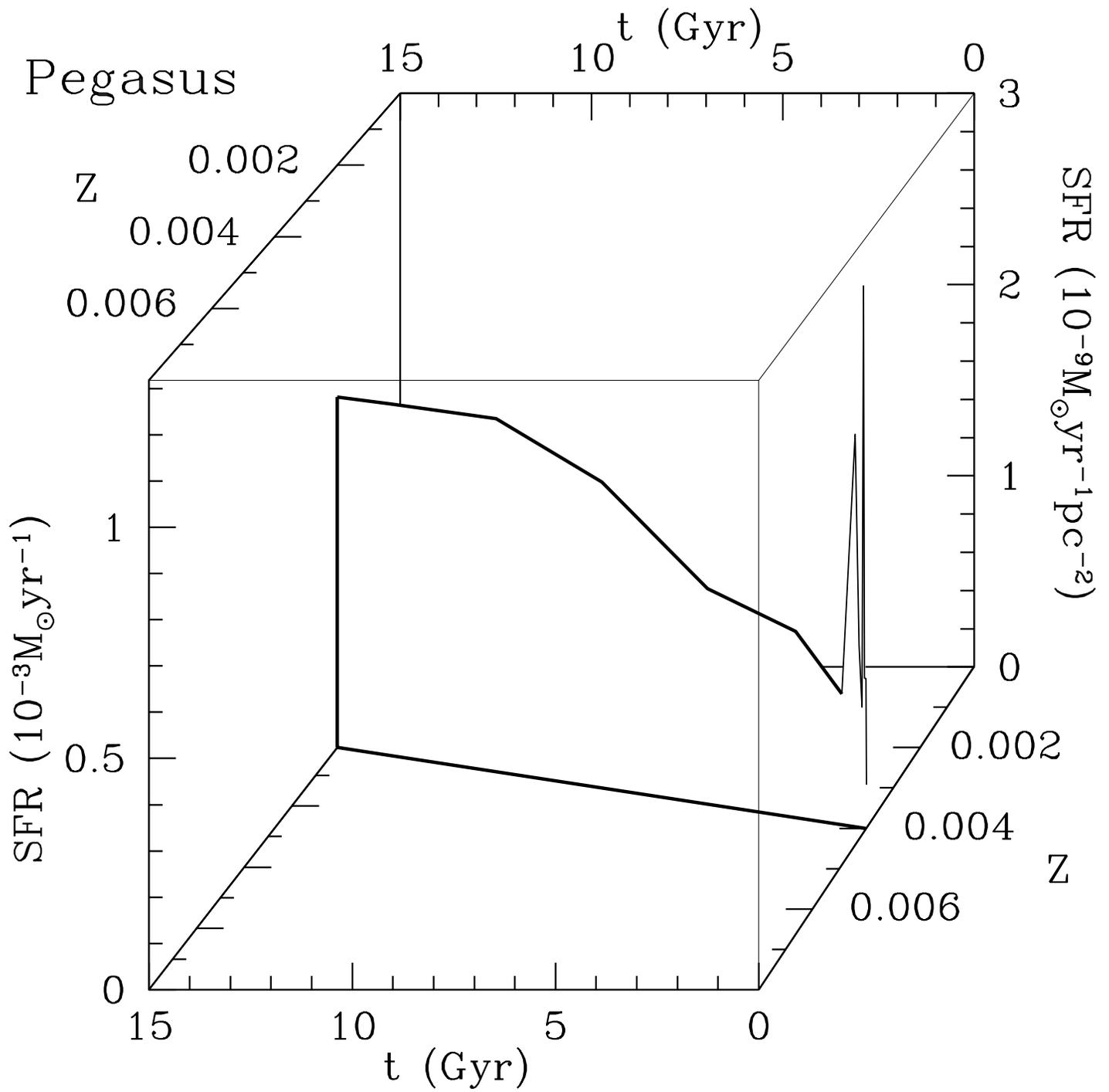
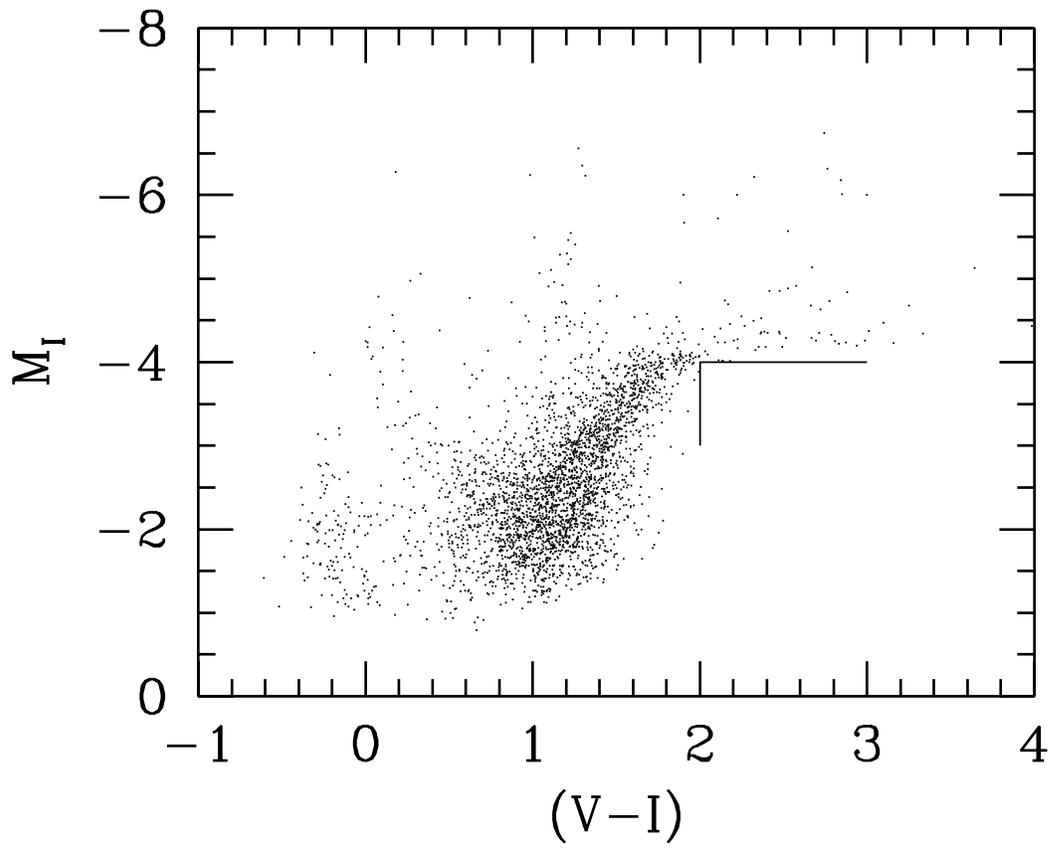
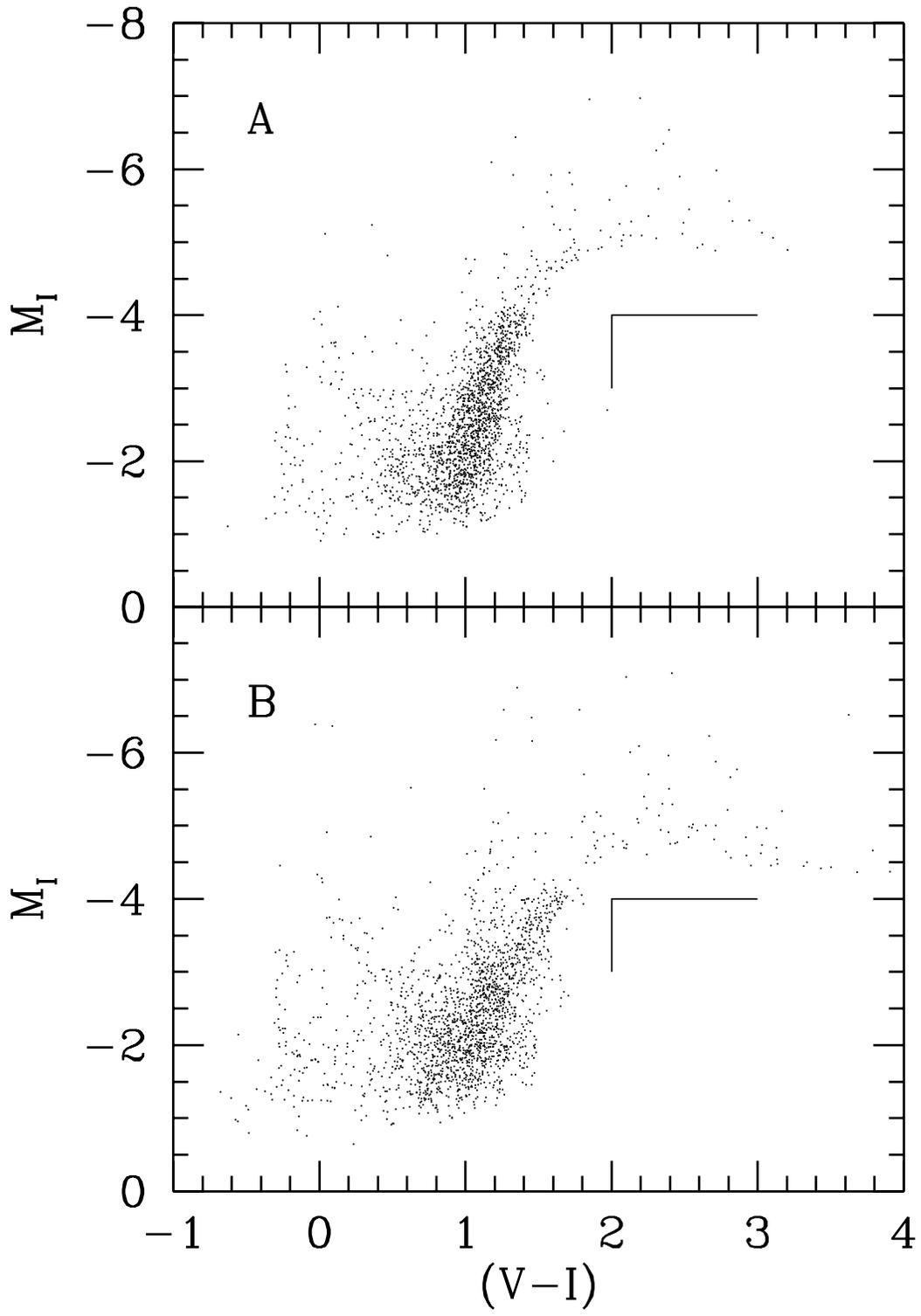
\end{document}